\documentclass[aps,preprint]{revtex4}%
\usepackage{amsfonts}
\usepackage{amsmath}
\usepackage{amssymb}
\usepackage{graphicx}%
\begin{document}
\title{Finite-Size Effects and Critical Behavior of the Deconfinement Phase Transition}
\author{M. Ladrem, A. Ait-El-Djoudi and G. Yezza}
\affiliation{Laboratoire de Physique des Particules et Physique Statistique. Department of
Physics, Ecole Normale Sup\'{e}rieure-Kouba, BP92, Vieux-Kouba, 16050,
Algiers, Algeria.}

\begin{abstract}
We study the finite-size effects on the deconfinement phase transition (DPT)
of hot and / or dense hadronic matter, using a simple thermodynamic model
based on the assumption of coexistence of confined and deconfined phases in a
finite volume, with the two-phases matter Equations of State (EoS). For the
QGP, we consider a partition function (PF) with the exact color-singletness
requirement. A problem arises in the limit of small QGP volumes when using the
usual color-singlet partition function (CSPF) derived in the saddle point
approximation. To avoid this problem, we have then proposed a method for
calculating a suitable CSPF which allows us to accurately calculate physical
quantities describing well the DPT at finite volumes like the order parameter,
the energy density and the entropy density. We show that in the limit of
infinite volume, these thermodynamic quantities exhibit a discontinuity at a
critical temperature $T_{c}(\infty)$ if the transition is temperature driven,
or chemical potential $\mu_{c}(\infty)$ if it is density driven. In a finite
size system, all singularities are smoothed out over a broadened critical
region, shifted from the critical point position in the thermodynamic limit.
At the level of the first derivatives of these thermodynamic quantities, the
specific heat and the susceptibility show delta function singularities at the
critical point in the thermodynamic limit. In a finite volume, these delta
singularities are smeared into finite peaks of widths $\delta T(V)$ or
$\delta\mu(V)$, with the maxima of the peaks occuring at pseudo-critical
points $T_{c}(V)$ or $\mu_{c}(V)$. For the temperature driven DPT, an analysis
of the finite size scaling behavior at criticality of these maxima as well as
of the width of the transition region and the shift of the transition
temperature allows us to determine the critical exponents characterizing the
deconfinement phase transition. The obtained results are in good agreement
with those predicted by other studies for a first-order phase transition.

\end{abstract}
\maketitle

\section{Introduction}

Phase transitions in statistical physics are known to be infinitely sharp only
in the thermodynamic limit $\left(  V\longrightarrow\infty\right)  $. Only in
this limit is the thermodynamical potential or any of its derivatives singular
at the critical point. However, real systems and the systems we simulate are
finite. For instance, the reaction zone in which the formation of a
quark-gluon plasma (QGP) in a relativistic heavy ion collision is assumed to
take place is limited, with a width of about $1\,fermi$, and a transverse area
of a radius comparable to the radii of the colliding nuclei, i. e., $V\simeq
A^{2/3}fm^{3}$. Also, lattice QCD calculations are necessarily performed in
finite lattices.

In general, finite size effects lead to a mixed phases system and a rounding
of the transition. The main care is then how to sign a possible phase
transition in a finite system. It turns out that we can extract the true
critical behavior of infinite systems from calculations on finite systems, by
studying how some characteristic thermodynamical quantities vary with the size
of the system, namely by a \textit{Finite Size Scaling} (FSS) analysis.
Studies in statistical physics have shown that despite the apparent diversity
in the underlying structure of systems undergoing phase transitions, these
latter take place with some \textit{universal} global behavior, depending only
on the range of interaction of the forces at play and the dimensionality of
the problem. This means the singular behavior of some characteristic
observables near criticality is identical for many systems when appropriately
scaled. This universal behavior of finite systems at criticality is
characterized by indices called \textit{critical exponents}. The determination
of critical exponents has long been one of the main interests for both
analytical calculations and numerical simulations.

In the present work, we study the finite size effects on the deconfinement
phase transition (DPT) when temperature driven and when density driven, within
a simple thermodynamical model used in \cite{SSG98}, based on the assumption
of the coexistence of confined and deconfined phases in a finite volume, and
using the equations of state (EoS) of the two phases. For the QGP, we consider
a partition function with the exact color-singletness requirement. This latter
is usually derived using the saddle point approximation, and gaussian
approximation which break down at $VT^{3}<<1$ \cite{EGR83,EGR84,EG86}. To
avoid the problem at small temperature and / or volumes, we have then proposed
a method for calculating a suitable CSPF, expanded in a power series of the
QGP volume fraction, which allows us to accurately calculate physical
quantities describing well the DPT at finite volumes like the order parameter,
the energy density and the entropy density. We show that in the limit of
infinite volume, these thermodynamic quantities exhibit a discontinuity at a
critical temperature $T_{c}(\infty)$ for a temperature driven DPT. In a finite
size system, no singularity occurs and the variations of these quantities are
perfectly smooth over a broadened critical region, shifted from the critical
point position in the thermodynamic limit.

At the level of the first derivatives of these thermodynamic quantities, the
specific heat and the susceptibility show delta function singularities at the
critical temperature in the thermodynamic limit. In finite volumes, these
delta singularities are smeared into finite peaks of widths $\delta T(V)$,
with the maxima of the peaks occuring at pseudo-critical temperatures
$T_{c}(V)$. An analysis of the finite size scaling behavior at criticality of
these maxima as well as of the width of the transition region and the shift of
the effective transition temperature $T_{c}(V)$ relative to the true one
$T_{c}(\infty)$ allows us to determine the critical exponents characterizing
the deconfinement phase transition.

\section{Color-singlet partition function of the QGP}

The partition function for a color-singlet quark-gluon plasma contained in a
volume $V_{QGP},$ at temperature $T$ and quark chemical potential $\mu,$ is
determined by \cite{EGR83,EGR84,EG86}:
\begin{equation}
Z(T,V_{QGP},\mu)=\frac{8}{3\pi^{2}}e^{-\frac{BV_{QGP}}{T}}\int_{-\pi}^{+\pi
}\int_{-\pi}^{+\pi}d\left(  \frac{\varphi}{2}\right)  d\left(  \frac{\psi}%
{3}\right)  M(\varphi,\psi)\widetilde{Z}(T,V_{QGP},\mu;\varphi,\psi),
\end{equation}
where the factor $\exp(-\frac{BV_{QGP}}{T})$\ accounts for the vacuum pressure
$B$, $M(\varphi,\psi)$ is the weight function (Haar measure) given by:
\begin{equation}
M(\varphi,\psi)=\left(  \sin\left(  \frac{1}{2}(\psi+\frac{\varphi}%
{2})\right)  \sin(\frac{\varphi}{2})\sin\left(  \frac{1}{2}(\psi-\frac
{\varphi}{2})\right)  \right)  ^{2},
\end{equation}
and $\widetilde{Z}$\ the generating function defined by:
\begin{equation}
\widetilde{Z}(T,V_{QGP},\mu;\varphi,\psi)=Tr\left[  \exp\left(  -\beta\left(
\widehat{H}_{0}-\mu\left(  \widehat{N}_{q}-\widehat{N}_{\overline{q}}\right)
\right)  +i\varphi\widehat{I}_{3}+i\psi\widehat{Y}_{8}\right)  \right]
\end{equation}
where $\beta=\dfrac{1}{T}$ with the units chosen as: $k_{B}=\hbar=c=1,$
$\widehat{H}_{0}$\ is the free quark-glue Hamiltonian, $\widehat{N}_{q}$
$\left(  \widehat{N}_{\overline{q}}\right)  $\ denotes the (anti-) quark
number operator, and $\widehat{I}_{3}$\ and $\widehat{Y}_{8}$\ are the color
``isospin'' and ``hypercharge'' operators respectively.

The generating function $\widetilde{Z}\left(  T,V_{QGP},\mu;\varphi
,\psi\right)  $\ can be factorized into the quark contribution and the glue
contribution as:
\begin{equation}
\widetilde{Z}\left(  T,V_{QGP},\mu;\varphi,\psi\right)  =\widetilde{Z}%
_{quark}\left(  T,V_{QGP},\mu;\varphi,\psi\right)  \widetilde{Z}_{glue}\left(
T,V_{QGP};\varphi,\psi\right)  ,
\end{equation}
where the quark contribution is given by:
\begin{equation}
\widetilde{Z}_{quark}\left(  T,V_{QGP},\mu;\varphi,\psi\right)  =\exp\left[
\frac{\pi^{2}}{12}T^{3}V_{QGP}d_{Q}\sum\limits_{q=r,g,b}\left(  \frac{7}%
{30}-\left(  \frac{\alpha_{q}-i(\frac{\mu}{T})}{\pi}\right)  ^{2}+\frac{1}%
{2}\left(  \frac{\alpha_{q}-i(\frac{\mu}{T})}{\pi}\right)  ^{4}\right)
\right]  ,
\end{equation}
with $q=r,\,b,\,g$ the color indices, $d_{Q}=2N_{f}$\ counts the spin-isospin
degeneracy of quarks\ and the angles$\ \alpha_{q}$ being determined by the
eigenvalues of the color charge operators in eq. (3):
\begin{equation}
\alpha_{r}=\frac{\varphi}{2}+\frac{\psi}{3},\;\;\alpha_{g}=-\frac{\varphi}%
{2}+\frac{\psi}{3},\;\;\alpha_{b}=-\frac{2\psi}{3},
\end{equation}
and the glue contribution is given by:
\begin{equation}
\widetilde{Z}_{glue}\left(  T,V_{QGP};\varphi,\psi\right)  =\exp\left[
\frac{\pi^{2}}{12}T^{3}V_{QGP}d_{G}\sum\limits_{g=1}^{4}\left(  -\frac{7}%
{30}+\left(  \frac{\alpha_{g}-\pi}{\pi}\right)  ^{2}-\frac{1}{2}\left(
\frac{\alpha_{g}-\pi}{\pi}\right)  ^{4}\right)  \right]  ,
\end{equation}
with $d_{G}=2$\ the degeneracy factor of gluons and $\alpha_{g}$ $\left(
g=1,...4\right)  $\ being:
\begin{equation}
\alpha_{1}=\alpha_{r}-\alpha_{g},\;\;\alpha_{2}=\alpha_{g}-\alpha
_{b},\;\;\alpha_{3}=\alpha_{b}-\alpha_{r},\;\;\alpha_{4}=0.
\end{equation}
Let us write the generating function as:
\begin{equation}
\widetilde{Z}\left(  T,V_{QGP},\mu;\varphi,\psi\right)  =e^{V_{QGP}%
T^{3}g(\varphi,\psi,\frac{\mu}{T})},
\end{equation}
with:
\begin{align}
g(\varphi,\psi,\frac{\mu}{T})  &  =\frac{\pi^{2}}{12}(\frac{21}{30}d_{Q}%
+\frac{16}{15}d_{G})+\frac{\pi^{2}}{12}\frac{d_{Q}}{2}\sum_{q=r,b,g}\left\{
-1+\left(  \frac{\left(  \alpha_{q}-i(\frac{\mu}{T})\right)  ^{2}}{\pi^{2}%
}-1\right)  ^{2}\right\} \nonumber\\
&  -\frac{\pi^{2}}{12}\frac{d_{G}}{2}\sum_{g=1}^{4}\left(  \frac{\left(
\alpha_{g}-\pi\right)  ^{2}}{\pi^{2}}-1\right)  ^{2},
\end{align}
then eq. (1) can be written on the form:
\begin{equation}
Z(T,\mu,V_{QGP})=\dfrac{8}{3\pi^{2}}e^{xV\left(  T^{3}g_{0}(\frac{\mu}%
{T})-\frac{B}{T}\right)  }\int_{-\pi}^{+\pi}\int_{-\pi}^{+\pi}d(\varphi
/2)d(\psi/3)M(\varphi,\psi)e^{(g(\varphi,\psi,\frac{\mu}{T})-g_{0}(\frac{\mu
}{T}))xVT^{3}},
\end{equation}
where $g_{0}\left(  \frac{\mu}{T}\right)  $\ is the maximum of $g(\varphi
,\psi,\frac{\mu}{T})$\ for $\varphi,\psi\in\left[  -\pi,+\pi\right]  $, given
as:
\begin{equation}
g_{0}\left(  \frac{\mu}{T}\right)  =\dfrac{\pi^{2}}{12}(\dfrac{21}{30}%
d_{Q}+\dfrac{16}{15}d_{G})+\dfrac{\pi^{2}d_{Q}}{12}(\dfrac{3\mu^{2}}{\pi
^{2}T^{2}}+\dfrac{3\mu^{4}}{2\pi^{4}T^{4}}),
\end{equation}
and $x$\ the QGP volume fraction defined by: $x=1-\mathfrak{h}$, since
assuming the phases coexistence model, the hadronic gas and QGP phases occupy
fractional volumes $V_{HG}=\mathfrak{h}V$ and $V_{QGP}=\left(  1-\mathfrak{h}%
\right)  V,$ characterized by the parameter $\mathfrak{h}$\ lying in the range
$0-1$, where $\mathfrak{h}=1$ corresponds to a pure hadron phase and
$\mathfrak{h}=0$ to a pure QGP phase.

When $VT^{3}\gg1,$ we can evaluate the integral in eq. (11) by the saddle
point approximation around the maximum of the integrand at $(0,0).$ After some
calculation, eq. (11) can be put on the form \cite{YG02}:
\begin{equation}
Z(T,\mu,Vx)\approx\dfrac{4}{9\pi^{2}}\dfrac{e^{xV\left(  T^{3}g_{0}(\frac{\mu
}{T})-\frac{B}{T}\right)  }}{\left(  VT^{3}a\left(  \frac{\mu}{T}\right)
\right)  ^{4}}\int_{-\pi}^{+\pi}\int_{-\pi}^{+\pi}d\varphi d\psi
M^{(0,0)}(\varphi,\psi)e^{xg^{(0,0)}(\varphi,\psi;VT^{3},\frac{\mu}{T})},
\end{equation}
with: $a\left(  \frac{\mu}{T}\right)  =\left(  \frac{d_{Q}}{16}\left(
\frac{3\mu^{2}}{\pi^{2}T^{2}}+1\right)  +\frac{3}{8}d_{G}\right)  ,$
$M^{(0,0)}$\ and $g^{(0,0)}$\ the expansions of $M$ and $\left(
g-g_{0}\right)  $ respectively, around $\left(  \varphi,\psi\right)  =(0,0)$,
given by:
\begin{equation}
\left\{
\begin{array}
[c]{c}%
M^{(0,0)}(\varphi,\psi)=\dfrac{1}{64}\varphi^{2}\left(  \psi^{2}%
-\dfrac{\varphi^{2}}{4}\right)  ^{2}\qquad
\;\;\;\;\;\;\;\;\;\;\;\;\;\;\;\;\;\;\;\;\;\;\;\;\;\;\;\;\;\;\;\;\;\;\;\;\;\;\;\;\;\;\;\;\;\;\;\;\;\;\;\;\;\;\;\;\;\\
g^{(0,0)}(\varphi,\psi;VT^{3},\frac{\mu}{T})=-\dfrac{2}{3}\left(  \varphi
^{2}+\dfrac{4}{3}\psi^{2}\right)  +\dfrac{1}{\pi\left(  a\left(  \frac{\mu}%
{T}\right)  \right)  ^{3/2}\sqrt{VT^{3}}}\left(  \dfrac{\varphi^{3}}%
{4}-\varphi\psi^{2}\right) \\
-\dfrac{7}{12\pi^{2}\left(  a\left(  \frac{\mu}{T}\right)  \right)  ^{2}%
VT^{3}}\left(  \dfrac{\varphi^{4}}{8}+\dfrac{2\psi^{4}}{9}+\dfrac{\varphi
^{2}\psi^{2}}{3}\right)  .
\end{array}
\right.
\end{equation}

Let us write the exponential in the integral of eq. (13) for $VT^{3}\gg
1$\ as:
\begin{align}
e^{xg^{(0,0)}(\varphi,\psi;VT^{3},\frac{\mu}{T})}  &  =e^{-\frac{2}{3}\left(
\varphi^{2}+\frac{4}{3}\psi^{2}\right)  x}\left(  1+\dfrac{x}{\pi\left(
a\left(  \frac{\mu}{T}\right)  \right)  ^{3/2}\sqrt{VT^{3}}}\left(
\dfrac{\varphi^{3}}{4}-\varphi\psi^{2}\right)  \right. \nonumber\\
&  \left.  -\dfrac{7x}{12\pi^{2}\left(  a\left(  \frac{\mu}{T}\right)
\right)  ^{2}VT^{3}}\left(  \dfrac{\varphi^{4}}{8}+\dfrac{2\psi^{4}}{9}%
+\dfrac{\varphi^{2}\psi^{2}}{3}\right)  \right)  ,
\end{align}
and do the expansion:
\begin{equation}
e^{-\frac{2}{3}\left(  \varphi^{2}+\frac{4}{3}\psi^{2}\right)  x}=\sum
_{j=0}^{\infty}\frac{\left(  -\frac{2}{3}\left(  \varphi^{2}+\frac{4}{3}%
\psi^{2}\right)  \right)  ^{j}}{j!}x^{j};
\end{equation}
then we obtain for the CSPF of the QGP:
\begin{equation}
Z_{QGP}(T,\mu,Vx)=\dfrac{4}{9\pi^{2}}\dfrac{e^{xV\left(  T^{3}g_{0}(\frac{\mu
}{T})-\frac{B}{T}\right)  }}{\left(  a\left(  \frac{\mu}{T}\right)
VT^{3}\right)  ^{4}}\left(  \sum_{j=0}^{\infty}\alpha_{j}x^{j}-\dfrac{7}%
{12\pi^{2}\left(  a\left(  \frac{\mu}{T}\right)  \right)  ^{2}VT^{3}}%
\sum_{j=0}^{\infty}\beta_{j}x^{j+1}\right)  ,
\end{equation}
such as the coefficients $\alpha_{j}$ and $\beta_{j}$\ of the two series are
given by:
\begin{equation}
\left\{
\begin{array}
[c]{c}%
\alpha_{j}=\int_{-\pi}^{\pi}\int_{-\pi}^{\pi}d\varphi d\psi M^{(0,0)}%
(\varphi,\psi)\dfrac{\left(  -\frac{2}{3}\left(  \varphi^{2}+\frac{4}{3}%
\psi^{2}\right)  \right)  ^{j}}{j!}\qquad\qquad\qquad\qquad\qquad\;\;\;\\
\beta_{j}=\int_{-\pi}^{\pi}\int_{-\pi}^{\pi}d\varphi d\psi M^{(0,0)}%
(\varphi,\psi)\left(  \dfrac{\varphi^{4}}{8}+\dfrac{2\psi^{4}}{9}%
+\dfrac{\varphi^{2}\psi^{2}}{3}\right)  \dfrac{\left(  -\frac{2}{3}\left(
\varphi^{2}+\frac{4}{3}\psi^{2}\right)  \right)  ^{j}}{j!}.
\end{array}
\right.
\end{equation}

It can easily be shown that the series in eq. (17) are convergent, and can be
truncated at a rank $j$ above which the contribution of the terms $\alpha
_{j}x^{j}$\ to the sums is negligible.

\section{Finite size effects}

To study the effects of volume finiteness on the DPT within the phases
coexistence model (PCM), let us examine the behavior of some thermodynamic
quantities in the vicinity of the critical point. The quantities of interest
are the order parameter, which is simply in this case the hadronic volume
fraction, the energy density and the entropy density. The mean value of any
intensive thermodynamic quantity of the system can be calculated by:
\begin{equation}
\langle A(T,\mu,V)\rangle=\frac{\int\limits_{0}^{1}\left(  \left(
1-\mathfrak{h}\right)  A_{Q}\left(  (1-\mathfrak{h})V\right)  +\mathfrak{h}%
\,A_{H}\left(  \mathfrak{h}V\right)  \right)  \,Z\left(  \mathfrak{h}\right)
\,d\mathfrak{h}}{\int\limits_{0}^{1}Z\left(  \mathfrak{h}\right)
\,d\mathfrak{h}},
\end{equation}
where $A_{QGP}\left(  \mathfrak{h}\right)  $ and $A_{HG}\left(  \mathfrak{h}%
\right)  $ are contributions relative to the individual QGP and hadronic gas
phases, respectively, and $Z\left(  \mathfrak{h}\right)  $\ the total PF of
the mixed system, which factorizes into the QGP and hadronic gas partition
functions as: $Z\left(  \mathfrak{h}\right)  =Z_{QGP}\left(  \mathfrak{h}%
\right)  Z_{HG}\left(  \mathfrak{h}\right)  .$

For the hadronic gas phase, let us consider the partition function of a pion
gas given by:
\begin{equation}
Z_{HG}=e^{-\frac{\pi^{2}}{30}T^{3}V_{HG}}\;.
\end{equation}

The mean value of the order parameter is then given by:
\begin{equation}
<\mathfrak{h}(T,\mu,V)>=1-\frac{\int\limits_{0}^{1}dx\exp\left(  \left(
(\frac{\pi^{2}}{30}+g_{0}(\frac{\mu}{T}))T^{3}-\frac{B}{T}\right)  xV\right)
\left(  \sum_{j=0}^{\infty}\alpha_{j}x^{j+1}-\dfrac{7\sum_{j=0}^{\infty}%
\beta_{j}x^{j+2}}{12\pi^{2}\left(  a\left(  \frac{\mu}{T}\right)  \right)
^{2}VT^{3}}\right)  }{\int\limits_{0}^{1}dx\exp\left(  \left(  (\frac{\pi^{2}%
}{30}+g_{0}(\frac{\mu}{T}))T^{3}-\frac{B}{T}\right)  xV\right)  \left(
\sum_{j=0}^{\infty}\alpha_{j}x^{j}-\dfrac{7\sum_{j=0}^{\infty}\beta_{j}%
x^{j+1}}{12\pi^{2}\left(  a\left(  \frac{\mu}{T}\right)  \right)  ^{2}VT^{3}%
}\right)  },
\end{equation}
and those of the energy and entropy densities can be written, respectively,
as:
\begin{align}
&  <\varepsilon(T,\mu,V)>=e_{HG}+e_{0}+\left(  e_{QGP}-e_{HG}\right)
<x(T,\mu,V)>\\
&  <s(T,\mu,V)>=s_{HG}+s_{0}+\left(  s_{QGP}-s_{HG}\right)  <x(T,\mu,V)>,
\end{align}
with:
\begin{equation}
\left\{
\begin{array}
[c]{c}%
<x(T,\mu,V)>=1-\,<\mathfrak{h}(T,\mu,V)>\hspace{6.7cm}\\
e_{HG}=\frac{\pi^{2}}{10}T^{4},\;\;e_{0}=-12\frac{T}{V},\;\;e_{QGP}=\frac
{\pi^{2}}{4}\left(  \frac{7N_{f}}{5}+\frac{32}{15}\right)  T^{4}-\frac{N_{f}%
}{4\pi^{2}}\mu^{4}+\frac{N_{f}}{2}T^{2}\mu^{2}+B\;\;\;\;\;\;\\
s_{HG}=\frac{2\pi^{2}}{15}T^{3},\;\;s_{0}=-\frac{12}{V}-\frac{4}{V}\ln\left(
VT^{3}a\left(  \frac{\mu}{T}\right)  \right)  ,\;\;s_{QGP}=\frac{\pi^{2}}%
{3}\left(  \frac{7N_{f}}{5}+\frac{32}{15}\right)  T^{3}+N_{f}\mu^{2}T.
\end{array}
\right.
\end{equation}

As it is known, the deconfined quark matter state can be obtained at extreme
conditions of temperature and/or density, i. e., either by raising temperature
or chemical potential. The deconfinement phase transition is then temperature
driven or density driven.

We'll examine in the following the FSE for a temperature driven DPT, at a
vanishing chemical potential $\left(  \mu=0\right)  $, and for a density
driven DPT at a fixed temperature $\left(  T=100MeV\right)  ,$\ considering
the two lightest quarks $u$\ and $d$ $\left(  N_{f}=2\right)  $, and using
$B^{1/4}=145MeV$\ for the bag constant.

Fig.(1-a) illustrates the variations of the order parameter, the energy
density normalized by $T^{4}$,\ and the entropy density normalized by $T^{3}$,
versus temperature for different system sizes. The curves show that in the
limit of an infinite volume, the three quantities exhibit a sharp
discontinuity at a transition temperature $T_{c}\left(  \infty\right)  $,
reflecting the first order character of the transition. For small size
systems, the transition is smoothed out over a range of temperature $\delta
T\left(  V\right)  $, around a pseudo-critical temperature $T_{c}\left(
V\right)  $, shifted from the true transition temperature $T_{c}\left(
\infty\right)  $. The broadening of the transition region, as well as the
shift of the critical temperature get larger, smaller is the volume.

In the same way as for temperature, we can deal with the density driven phase
transition, at a fixed temperature $\left(  T=100MeV\right)  $. The variations
of the order parameter, the energy density\ and the entropy density versus
chemical potential are presented for various system sizes on Fig. (1-b). The
first-order character of the transition can, also in this case, clearly be
seen at the large volumes limit. In small systems, the transition is perfectly
smooth over a region of chemical potential of width $\delta\mu\left(
V\right)  $, and the effective critical chemical potential $\mu_{c}\left(
V\right)  $\ is shifted away from the true one $\mu_{c}\left(  \infty\right)
.$

\section{Finite-size scaling analysis}

\subsection{Finite size scaling and critical exponents}

In the thermodynamic limit, phase transitions are characterized by the
appearance of singularities in some second derivatives of the thermodynamic
potential, such as the susceptibility $\chi$ and the specific heat $c$. For a
first order phase transition, the divergences are typically $\delta$-function
singularities, corresponding to the discontinuities in the first derivatives
of the thermodynamic potential.

For temperature driven phase transitions, in finite volumes, the singularities
in $\chi\left(  T,V\right)  $ and $c\left(  T,V\right)  $\ are rounded over a
range of temperature $\delta T\left(  V\right)  .$\ The peaks occuring at a
pseudo-critical temperature $T_{c}\left(  V\right)  $\ may be shifted away
from the true critical temperature $T_{c}\left(  \infty\right)  $. The width
of the transition region, the shift of the critical temperature and the maxima
of the peaks of the specific heat and the susceptibility, represented in Fig.
(2), show a scaling behavior at criticality, characterized by critical
exponents as:
\begin{equation}
\left\{
\begin{array}
[c]{c}%
\delta T\left(  V\right)  \sim V^{-\theta_{T}}%
\;\;\;\;\;\;\;\;\;\;\;\;\;\;\;\;\;\;\;\;\;\;\;\;\;\;\;\;\;\\
\tau_{T}(V)=T_{c}\left(  V\right)  -T_{c}\left(  \infty\right)  \sim
V^{-\lambda_{T}}\\
c_{T}^{\max}\left(  V\right)  \sim V^{\alpha_{T}}%
\;\;\;\;\;\;\;\;\;\;\;\;\;\;\;\;\;\;\;\;\;\;\;\;\\
\chi_{T}^{\max}\left(  V\right)  \sim V^{\gamma_{T}}%
.\;\;\;\;\;\;\;\;\;\;\;\;\;\;\;\;\;\;\;\;\;\;\;
\end{array}
\right.
\end{equation}

The finite size scaling (FSS) analysis is used to recover the critical
exponents, and it has been shown \cite{BH88} that for a first order phase
transition , the finite size quantities $\delta T\left(  V\right)  $,
$\tau_{T}(V)$ scale as: $V^{-1}$\ and the maxima $c_{T}^{\max}\left(
V\right)  $\ and $\chi_{T}^{\max}\left(  V\right)  $\ scale as: $V$. The
critical exponents $\theta_{T}$, $\lambda_{T}$, $\alpha_{T}$ and $\gamma_{T}%
$\ are then all equal to the dimensionality.

\subsection{Numerical determination of the critical exponents}

\subsubsection{\textit{The susceptibility critical exponent}}

For the temperature driven phase transition, the susceptibility is defined to
be the first derivative of the order parameter\ with respect to temperature,
i. e.,
\begin{equation}
\chi\left(  T,V\right)  =\dfrac{\partial\langle\mathfrak{h}\left(  T,V\right)
\rangle}{\partial T}.
\end{equation}
It is clear in this case from Fig. (3-left) that the delta function
singularity of the susceptibility in the thermodynamic limit is smeared, in a
finite volume, into a finite peak of width $\delta T\left(  V\right)  $, with
the maximum of the peak $\mid\chi_{T}\mid^{\max}\left(  V\right)  $ occuring
at the pseudo-critical temperature $T_{c}(V)$. The plot of the maxima
$\mid\chi_{T}\mid^{\max}\left(  V\right)  $\ versus volume is illustrated in
Fig. (3-right), and the linearity of the data with $V$ can clearly be noted. A
numerical parametrization with the power-law form: $\mid\chi_{T}\mid^{\max
}\left(  V\right)  \sim V^{\gamma_{T}},$ gives the value\ of the
susceptibility critical exponent $\gamma_{T}.$ The determination of the
location of the maxima being done using a numerical method with an error on
the temperature and on the maxima $\mid\chi_{T}\mid^{\max}\left(  V\right)  $,
this yields a systematic error on the exponent $\gamma_{T}.$ The systematic
error is given for all the critical exponents determined in the following. The
result for the susceptibility critical exponent and the estimated systematic
error are then: $\gamma_{T}=0.99\pm0.04$.

\subsubsection{\textit{The specific heat critical exponent}}

The specific heat density representing the smeared delta function of the
latent heat is defined as:
\begin{equation}
c\left(  T,V\right)  =\dfrac{\partial\langle\varepsilon\left(  T,V\right)
\rangle}{\partial T},
\end{equation}
$\varepsilon$ being the energy density of the system. The variations of the
specific heat density with temperature are presented for different sizes in
Fig. (4-left), which shows the rounding of the delta function singularity of
$c\left(  T,V\right)  $ in finite systems into a finite peak of width $\delta
T\left(  V\right)  $\ and height $c_{T}^{\max}\left(  V\right)  .$ For
decreasing volume, the width get larger while the height of the peak
decreases. The data of the maxima of the specific heat $c_{T}^{\max}\left(
V\right)  $\ are fitted to the power-law form: $c_{T}^{\max}\left(  V\right)
\sim V^{\alpha_{T}}$ in Fig. (4-right),\ and the obtained specific heat
critical exponent is: $\alpha_{T}=0.99\pm0.04.$

\subsubsection{\textit{The shift critical exponent}}

For the study of the shift of the transition temperature $\tau_{T}%
(V)=T_{c}\left(  V\right)  -T_{c}\left(  \infty\right)  $, we need to have the
effective transition temperature\ in a finite volume. It is usually defined to
be the temperature at which the rounded peaks of the susceptibility and the
specific heat reach their maxima$.$

Fig. (5) illustrates the results for the shift of the critical temperature
plotted versus inversed volume, and shows the linear character of the
variations. The shift critical exponent obtained from a fit to the form:
$\tau_{T}(V)\sim V^{-\lambda_{T}}$, is: $\lambda_{T}=1.0085\pm0.0009.$

\subsubsection{\textit{The smearing critical exponent}}

The width of the transition region can be defined by the gap: $\delta
T(V)=T_{2}(V)-T_{1}(V)$ with $T_{1}(V)$ and $T_{2}(V)$ the temperatures at
which the second derivative of the order parameter reaches its maxima, or in
other terms the temperatures at which the third derivative of the order
parameter vanishes, i. e.,
\begin{equation}
\left.  \frac{\partial^{3}}{\partial T^{3}}\langle\mathfrak{h}\left(
T,V\right)  \rangle\right|  _{T_{1}(V),T_{2}(V)}=0.
\end{equation}

Fig. (6-left) illustrates the variations of the second derivative of the order
parameter with temperature for various sizes, and shows that the gap between
the two extrema, which represents the broadening of the transition region,
decreases with increasing volume. The results for the widths $\delta T\left(
V\right)  $, plotted in Fig. (6-right) vs volume, were fitted to the power law
form: $\delta T\left(  V\right)  \sim V^{-\theta_{T}}$, and the obtained
smearing critical exponent is: $\theta_{T}=1.016\pm0.007.$

\section{Conclusion}

Our work has shown the influence of the finiteness of the system size on the
behavior of thermodynamical quantities near criticality. The sharp transition
observed in the thermodynamic limit, signaled by discontinuities in the first
derivatives of the thermodynamic potential at a critical temperature or
chemical potential, is rounded off in finite volume, and the variations of the
thermodynamic quantities are perfectly smooth on the hole range of temperature
or chemical potential. For the temperature driven DPT, a finite-size scaling
analysis of the behavior of the width of the transition region$\ \delta
T\left(  V\right)  $, the shift of the pseudo-critical temperature relative to
the true one $\tau_{T}\left(  V\right)  =T_{c}\left(  V\right)  -T_{c}\left(
\infty\right)  $, and the maxima of the rounded peaks of the susceptibility
$\chi_{T}^{\max}\left(  V\right)  $\ and the specific heat\ $c_{T}^{\max
}\left(  V\right)  $ near criticality, shows their power-law variations with
the volume:
\[
\left\{
\begin{array}
[c]{c}%
\delta T\left(  V\right)  \sim V^{-\theta_{T}}%
\;\;\;\;\;\;\;\;\;\;\;\;\;\;\;\;\;\;\;\;\;\;\;\;\;\;\\
\tau_{T}(V)=T_{c}\left(  V\right)  -T_{c}\left(  \infty\right)  \sim
V^{-\lambda_{T}}\\
c_{T}^{\max}\left(  V\right)  \sim V^{\alpha_{T}}%
\;\;\;\;\;\;\;\;\;\;\;\;\;\;\;\;\;\;\;\;\;\;\;\;\;\\
\chi_{T}^{\max}\left(  V\right)  \sim V^{\gamma_{T}}%
\;\;\;\;\;\;\;\;\;\;\;\;\;\;\;\;\;\;\;\;\;\;\;\;
\end{array}
\right.  \,,
\]
characterized by the scaling critical exponents $\theta_{T},\,\lambda
_{T},\,\alpha_{T}$ and $\gamma_{T}$. Numerical results for these critical
exponents have been obtained, and the associated systematic errors, resulting
from the numerical method of calculating the maxima $\chi_{T}^{\max}\left(
V\right)  $\ and $c_{T}^{\max}\left(  V\right)  $ and their localizations as
well as the gap $\delta T(V)=T_{2}(V)-T_{1}(V)$, have been estimated. Our
results for the critical exponents are in good agreement with the analytical
ones: $\theta_{T}=\lambda_{T}=\alpha_{T}=\gamma_{T}=1$ found in a parallel
work \cite{LYA02}. These results are characteristic of the first order phase
transition, as predicted by the FSS theory.

A further FSS analysis is viewed using an exact color-singlet partition
function of the QGP, derived without any approximation, with an appropriate
numerical treatment. Also, the density driven DPT can be studied, and similar
critical exponents associated to the shift of the critical chemical potential
$\left[  \mu_{c}\left(  V\right)  -\mu_{c}\left(  \infty\right)  \right]  $,
the width of the transition region $\delta\mu\left(  V\right)  $ and the
maxima of the rounded peaks of the susceptibility $\chi_{\mu}^{\max}\left(
V\right)  $ and the specific heat $c_{\mu}^{\max}\left(  V\right)  $ (relative
to the derivatives with respect to $\mu$) can be determined.

\end{document}